\documentstyle[12pt]{article}
\def\l{\label}
\def\pmb#1{\setbox0=\hbox{$#1$}%
  \kern-.025em\copy0\kern-\wd0
  \kern.05em\copy0\kern-\wd0
  \kern-.025em\raise.0433em\box0}
\def\parb{\pmb{\partial}}

\newcommand{\be}{\begin{equation}}
\newcommand{\ee}{\end{equation}} 
\newcommand{\eei}{\end{equation}\indent\indent}
\newcommand{\bc}{\begin{center}}
\newcommand{\ec}{\end{center}}
\newcommand{\bea}{\begin{eqnarray}}
\newcommand{\eea}{\end{eqnarray}}
\newcommand{\ba}{\begin{array}}
\newcommand{\ea}{\end{array}}

\newcommand{\sfrac}[2]{{\textstyle{#1\over#2}}}
\def\case#1/#2{\textstyle\frac{#1}{#2} }

\catcode`\@=11
\long\def\@makefntext#1{
\protect\noindent \hbox to 3.2pt {\hskip-.9pt
$^{{\ninerm\@thefnmark}}$\hfil}#1\hfill}		

 \def\@makefnmark{\hbox to 0pt{$^{\@thefnmark}$\hss}}  

\def\ps@myheadings{\let\@mkboth\@gobbletwo
\def\@oddhead{\hbox{}
\rightmark\hfil\ninerm\thepage}
\def\@oddfoot{}\def\@evenhead{\ninerm\thepage\hfil
\leftmark\hbox{}}\def\@evenfoot{}
\def\sectionmark##1{}\def\subsectionmark##1{}}

\newcounter{sectionc}\newcounter{subsectionc}\newcounter{subsubsectionc}
\renewcommand{\section}[1] {\vspace{0.6cm}\addtocounter{sectionc}{1}
\setcounter{subsectionc}{0}\setcounter{subsubsectionc}{0}\noindent
 	{\bf\thesectionc. #1}\par\vspace{0.4cm}}
\renewcommand{\subsection}[1] {\vspace{0.6cm}\addtocounter{subsectionc}{1}
 	\setcounter{subsubsectionc}{0}\noindent
 	{\it\thesectionc.\thesubsectionc. #1}\par\vspace{0.4cm}}
\renewcommand{\subsubsection}[1] {\vspace{0.6cm}\addtocounter{subsubsectionc}{1}
  	\noindent {\rm\thesectionc.\thesubsectionc.\thesubsubsectionc.
  	#1}\par\vspace{0.4cm}}

\newcounter{appendixc}
\newcounter{subappendixc}[appendixc]
\newcounter{subsubappendixc}[subappendixc]

\renewcommand{\appendix}[1] {\vspace{0.6cm}
        \refstepcounter{appendixc}
        \setcounter{figure}{0}
        \setcounter{table}{0}
        \setcounter{equation}{0}
        \renewcommand{\thefigure}{\Alph{appendixc}.\arabic{figure}}
        \renewcommand{\thetable}{\Alph{appendixc}.\arabic{table}}
        \renewcommand{\theappendixc}{\Alph{appendixc}}
        \renewcommand{\theequation}{\Alph{appendixc}.\arabic{equation}}
        \noindent{\bf Appendix \theappendixc #1}\par\vspace{0.4cm}}

\def\abstracts#1{{
 	\centering{\begin{minipage}{30pc}\tenrm\baselineskip=12pt\noindent
 	\centerline{\tenrm ABSTRACT}\vspace{0.3cm}
 	\parindent=0pt #1
 	\end{minipage}}\par}}

\renewenvironment{thebibliography}[1]
 	{\begin{list}{\arabic{enumi}.}
 	{\usecounter{enumi}\setlength{\parsep}{0pt}
\setlength{\leftmargin 1.25cm}{\rightmargin 0pt}
 	 \setlength{\itemsep}{0pt} \settowidth
 	{\labelwidth}{#1.}\sloppy}}{\end{list}}

\newcounter{itemlistc}
\newcounter{romanlistc}
\newcounter{alphlistc}
\newcounter{arabiclistc}

\newcommand{\fcaption}[1]{
        \refstepcounter{figure}
        \setbox\@tempboxa = \hbox{\tenrm Fig.~\thefigure. #1}
        \ifdim \wd\@tempboxa > 6in
           {\begin{center}
        \parbox{6in}{\tenrm\baselineskip=12pt Fig.~\thefigure. #1}
            \end{center}}
        \else
             {\begin{center}
             {\tenrm Fig.~\thefigure. #1}
              \end{center}}
        \fi}

\newcommand{\tcaption}[1]{
        \refstepcounter{table}
        \setbox\@tempboxa = \hbox{\tenrm Table~\thetable. #1}
        \ifdim \wd\@tempboxa > 6in
           {\begin{center}
        \parbox{6in}{\tenrm\baselineskip=12pt Table~\thetable. #1}
            \end{center}}
        \else
             {\begin{center}
             {\tenrm Table~\thetable. #1}
              \end{center}}
        \fi}

\def\@citex[#1]#2{\if@filesw\immediate\write\@auxout
 	{\string\citation{#2}}\fi
\def\@citea{}\@cite{\@for\@citeb:=#2\do
 	{\@citea\def\@citea{,}\@ifundefined
 	{b@\@citeb}{{\bf ?}\@warning
 	{Citation `\@citeb' on page \thepage \space undefined}}
 	{\csname b@\@citeb\endcsname}}}{#1}}

\newif\if@cghi
\def\cite{\@cghitrue\@ifnextchar [{\@tempswatrue
 	\@citex}{\@tempswafalse\@citex[]}}
\def\citelow{\@cghifalse\@ifnextchar [{\@tempswatrue
 	\@citex}{\@tempswafalse\@citex[]}}
\def\@cite#1#2{{$\null^{#1}$\if@tempswa\typeout
 	{IJCGA warning: optional citation argument
 	ignored: `#2'} \fi}}

\def\fnt#1#2{\footnotetext{\kern-.3em
 	{$^{\mbox{\sevenrm #1}}$}{#2}}}

 1
 1
 1

\font\tenbf=cmbx10
\font\tenrm=cmr10
\font\tenit=cmti10

\font\ninerm=cmr9

\begin{document}

\centerline{\tenbf SINGULARITIES IN SILENT UNIVERSES: STATE OF THE ART}
\vspace{0.8cm}
\centerline{\tenrm MARCO BRUNI}
\baselineskip=13pt
\centerline{\tenit 
ICTP, International Center for Theoretical Physics, P. O. Box 586,
34014 Trieste, Italy}
\centerline{\tenit SISSA, via Beirut 2-4, 34013 Trieste, Italy}
\vspace{0.3cm}
\centerline{\tenrm HENK VAN ELST }
\centerline{\tenit
Astronomy Unit, Queen Mary \& Westfield College,}
\centerline{\tenit
University of London, Mile End Road, London E1 4NS, UK}
\vspace{0.3cm}
\centerline{\tenrm CLAES UGGLA}
\centerline{\tenit Department of Physics, Stockholm
University, Box 6730, S--113 85 Stockholm, Sweden}
\centerline{\tenit and}
\centerline{\tenit Department of Physics, Lule\aa\ University of
Technology, S--951 87 Lule\aa, Sweden}

\vspace{0.9cm} 
\abstracts{ 
After a brief overview of the so-called
silent models and their present status, we consider the subclass of
Bianchi Type--I models with a magnetic field source. Due to the presence
of the magnetic field, the initial singularity shows ``oscillatory''
features reminiscent of the Bianchi Type--IX case. The Bianchi Type--I models
with a magnetic field are therefore a counterexample to the folklore
that matter fields can be neglected in the vicinity of the
singularity. }
\bigskip\bigskip\bigskip
\bigskip\bigskip\bigskip
\centerline{\it Proceedings of the }
\centerline{\it 12th Italian Conference on General
Relativity and Gravitational Physics}
\bigskip\bigskip\bigskip
\bigskip\bigskip\bigskip
\bigskip\bigskip\bigskip
\centerline{IC/96/223}
\centerline{SISSA--160/96/A}
\newpage
\rm\baselineskip=14pt
\section{Introduction}

Often a preferred timelike vector field $u^a$ exists in a cosmological
context (usually 
the  4-velocity of the fluid), and various kinematical and dynamical
quantities are defined with respect to $u^a$: for example, the
electric and magnetic parts of the Weyl tensor $C^a{}_{bcd}$ are,
respectively, $E^a{}_b:=C^a{}_{cbd}\,u^c\,u^d$ and
$H^a{}_b:=\sfrac{1}{2}\,\eta^a{}_{ce}{}^f\,C^e{}_{fbd}\,u^c\,u^d$.  
The so-called ``silent universes'' are cosmological models
characterized by two specific features: the first regards their matter content,
which is taken to be a perfect fluid with vanishing pressure, \hbox{$p=0$};
the second regards the dynamics, constrained by assuming the vanishing
of the magnetic part of the Weyl tensor, $H_{ab}=0$. 
The name ``silent'' itself
comes from noticing that the assumptions $p=H_{ab}=0$ imply 
the lack of sound and gravitational waves.\cite{bi:MPS2}
In the recent cosmological
literature, models characterized by $p=H_{ab}=0$ plus a third
assumption, the vanishing of the vorticity of the fluid
$\omega_{ab}=0$, were originally introduced by Matarrese {\it et al\/}
(1993) to model the non-linear evolution of perturbations during the
matter dominated era.\cite{bi:MPS1} It was hoped that these models
could give a reasonably good description of the growth of 
structure in a collisionless medium. The assumptions $p=\omega_{ab}=0$
are very standard in dealing with these problems; the restriction
$H_{ab}=0$ was introduced by extrapolating from the vanishing of $H_{ab}$
in first-order scalar perturbations.\cite{bi:BDE}  Matarrese {\it et
al} (1993) also implicitly assumed\cite{bi:MPS1} that the field equations were
compatible with the $H_{ab}=0$ constraint or, in other words, that the
evolution of initial data preserves $H_{ab}=0$. This assumption was
corroborated by a preceding paper by Barnes and Rowlingson\cite{bi:BR}
(1989), and later on by work\cite{bi:lesame} of Lesame {\it et al\/} (1995).

In studying models with $p=\omega_{ab}=H_{ab}=0$ one takes the
matter density $\mu$, the expansion scalar $\Theta$, the shear
$\sigma_{ab}$ and the electric Weyl tensor $E_{ab}$ as physical
variables; the dynamics of these quantities is then determined by a
set of field equations arising from the Ricci and the Bianchi
identities\footnote{In this approach, the Einstein equations are used
to establish an algebraic constraint between the Ricci part of the
curvature, and the energy-momentum tensor.}.  The
amazing feature of these models is that the evolution equations for
the variables $\mu$, $\Theta$, $\sigma_{ab}$, $E_{ab}$ become a set of
quasi-linear first-order ordinary differential equations (ODEs), as was
first explicitly shown\cite{bi:MPS1} by Matarrese {\it et al\/} (1993).
This fact itself was enough to raise the attention of cosmologists on
these silent models:\cite{bi:bert} would they provide a fair
description of the non-linear evolution of cosmological
perturbations? The assumptions $p=\omega_{ab}=H_{ab}=0$ would simplify
computer simulations of large-scale structure formation enormously. 
Unfortunately, it was soon realized by Matarrese {\it et
al\/}\cite{bi:MPS2} (1994) that the $H_{ab}=0$ assumption already breaks down
at the second order in the perturbations. In this context it should be
noted that the cosmological effects of $H_{ab}$ have never been properly
quantified. 
Assuming that the constraint $H_{ab}=0$
was compatible with the evolution equations for the physical
variables, the dynamics of silent models with $p=\omega_{ab}=H_{ab}=0$
was investigated from a
dynamical system point of view by Bruni {\it et al\/}
(1995), who  focused 
on the character of the singularities.\cite{bi:BMP1}  They found that the
generic final fate of a fluid element under the
$p=\omega_{ab}=H_{ab}=0$ restrictions is to fall into a spindle-like
Kasner singularity, while only a set of measure zero of initial data
leads to a pancake (which is instead thought to be generic in a Newtonian
collapse scenario). Finally, in a parallel work, Bruni {\it et al\/}
(1995) included a cosmological constant
$\Lambda$ in their analysis\cite{bi:BMP2}  
in order to investigate the so-called ``cosmic
no-hair" conjecture,\cite{bi:hair} and it was indeed found
that the de Sitter spacetime 
is the unique attractor for initial data leading to
expansion\footnote{It was also found\cite{bi:BMP2} that there are
also initial data for which, even after a period of inflation, a
recollapse occurs, in which case $\Lambda$ is soon negligible, and the
collapse leads to a Kasner spindle-like singularity, as in the
$\Lambda=0$ case. On the other hand, when $\Lambda=0$, the attractor
for initial data leading to expansion is the Milne model.\cite{bi:BMP1}}.
An important class of spatially
inhomogeneous exact solutions resides in the silent family, namely the
well-known Szekeres models.\cite{bi:szek}
These contain the spherically symmetric
Lema\^{\i}tre--Tolman--Bondi models which in turn have the spatially 
homogeneous Kantowski--Sachs models as a special subclass.
The Szekeres models are  of Petrov type D, and may be called
degenerate, in that the Weyl tensor (reducing to its electric part
$E_{ab}$) and the shear $\sigma_{ab}$ each have two equal eigenvalues.
Recently the existence of a more general spatially inhomogeneous
class of silent models (of non-degenerate Petrov type I) has been
challenged.\cite{bi:bonetal96,bi:vanetal96} Furthermore, even if
these models exist, their singularities  are unstable against $H_{ab}$
perturbations,\cite{bi:maedaetal} and their
relevance to structure formation is at least doubtful.\cite{bi:MPS2}
Therefore, in the following we shall focus on homogeneous spatially flat
silent models of Bianchi Type--I with a magnetic field, which do
exist as solutions of the Einstein--Maxwell field equations.

\section{Dynamics of spatially flat silent universes with a magnetic field}
As we said above, the evolution
equations one obtains under the assumptions
$p=\omega_{ab}=H_{ab}=0$ are ODEs. Then, it is natural to ask if these
restrictions can be
relaxed, or if other matter fields can be considered, such that the
ODE-structure of the evolution equations is retained. We have found that
this is indeed the case if
one has non-vanishing fluid vorticity $\omega_{ab}\not=0$, or if
one includes a magnetic field $H_a$ as an additional matter source. In the
latter case also the Maxwell equations come into play. When
$p=\omega_{ab}=H_{ab}=H_{a}=0$, the shear $\sigma_{ab}$ and the electric
Weyl tensor $E_{ab}$ admit a common Fermi-Walker transported
eigentetrad.\cite{bi:BR,bi:MPS1}
This is also true if a magnetic field is aligned with one of the shear
eigendirections. However, for non-zero fluid vorticity $\omega_{ab}$ or a
general magnetic field $H_a$, the eigentetrads of the shear and the electric
Weyl tensor are no longer Fermi--Walker transported, nor do they coincide.

One can show that non-shear-aligned magnetic fields are only compatible
with a spatially homogeneous Bianchi Type--I spacetime
geometry,\cite{bi:bruetal96} which we will consider in the following.
The kinematical quantities we use refer to the normal timelike
congruence associated with the spacelike surfaces of homogeneity. For
simplicity we take the energy-momentum tensor to be comprised of the
magnetic field $H_a$ only.\footnote{A dust matter field with $p_D=0$
and/or a cosmological $\Lambda$ can always be added, and indeed their
coming into play makes the discussion of the corresponding models more
interesting for physical purposes, but for the sake of simplicity we
shall not consider them here.} The Bianchi Type--I spacetime geometry
automatically implies $\omega_{ab} = H_{ab} = 0$.
Here, we adopt a tetrad description, where we denote the
spatial components of tensor fields by Greek indices, and we choose a
shear eigentetrad. 
{}From a dynamical point of view, the configuration outlined is completely
characterized by two of the shear eigenvalues, e.g., $\sigma_2$ and
$\sigma_3$, the expansion scalar $\Theta$, and the three magnetic field
component $H_\alpha$.  However,
in order to study their dynamical properties, it is very useful to introduce
five dimensionless variables, a dimensionless angular velocity
(Fermi-rotation), and a new time derivative:
\be
\Sigma_\pm:=\sigma_\pm/\Theta\;, ~~~
{\cal H}_\alpha:=\sqrt{3}H_\alpha/\Theta\;, ~~~
R_\alpha:=\Omega_\alpha/\Theta\;, ~~~
\parb_0:=3{\bf e}_0/\Theta\;,
\ee
where $\sigma_+:=3(\sigma_2+\sigma_3)/2$ and
$\sigma_-:=\sqrt{3}(\sigma_2-\sigma_3)/2$, $\Omega_{\alpha}$ gives the
angular velocity of the spatial tetrad with respect to a Fermi-transported
basis, and ${\bf e}_0$ denotes the time derivative along $u^a$. 
Finally, it is also useful to define the standard deceleration parameter
$q:=-3\dot{\Theta}/{\Theta}^2 -1$ and a density parameter
$\Omega_H:={\cal H}_\alpha{\cal H}^\alpha/2$, related to the shear
variables $\Sigma_\pm$ by
\be\l{eq:qom}
 q = 1+\Sigma_+{}^2+\Sigma_-{}^2;~~~ \Omega_H=1-\Sigma_+{}^2-\Sigma_-{}^2\;.
\ee
With these definitions, we obtain the following evolution equations:

\bea
\l{eq:dsigdotp1}
 \parb_{0} \Sigma_+ & = & - \,(2-q)\,\Sigma_+
 +{\cal H}_1{}^2 - \sfrac12 \,(\, {\cal H}_2{}^2 + {\cal H}_3{}^2\,)\;, \\ 
\l{eq:dsigdotm1}
 \parb_{0} \Sigma_- & = & - \,(2-q)\,\Sigma_-
- \sfrac{\sqrt{3}}{2}\, (\, {\cal H}_2{}^2 - {\cal H}_3{}^2\,)\;, \\
\l{eq:dmag1dot1}
\parb_{0}{\cal H}_{1} & = & - \,(1-q+2\,\Sigma_{+})\,
{\cal H}_{1} + 3\,(R_{2}\,{\cal H}_{3}-R_{3}\,{\cal H}_{2})\;,
\\ 
\l{eq:dmag2dot1}
\parb_{0}{\cal H}_{2} & = & - \,(1-q-\Sigma_{+}-\sqrt{3}\,
\Sigma_{-})\,{\cal H}_{2} + 3\,(R_{3}\,{\cal H}_{1}-R_{1}\,
{\cal H}_{3})\;, \\ 
\l{eq:dmag3dot1}
\parb_{0}{\cal H}_{3} & = & - \,(1-q-\Sigma_{+}+\sqrt{3}\,
\Sigma_{-})\,{\cal H}_{3} + 3\,(R_{1}\,{\cal H}_{2}-R_{2}\,
{\cal H}_{1}) \;,
\eea
where the $R_\alpha$ are determined by the algebraic constraints
%
%
\be\l{eq:alcon}
{\cal H}_1\,{\cal H}_2 + 3\,(\Sigma_2 -\Sigma_1)\,R_3=0\;,~~~
{\cal H}_2\,{\cal H}_3 + 3\,(\Sigma_3 -\Sigma_2)\,R_1=0\;,~~~
{\cal H}_3\,{\cal H}_1 + 3\,(\Sigma_1 -\Sigma_3)\,R_2=0\;,~~~
\ee
where $\Sigma_1=-\,2\,\Sigma_+/3$, $\Sigma_2=(\Sigma_++\sqrt{3}\,
\Sigma_-)/3$, $\Sigma_3=(\Sigma_+-\sqrt{3}\,\Sigma_-)/3$. 
The physical phase space of this dynamical system admits various invariant
submanifolds: the most important for the present discussion are the
three that arise when the magnetic field is aligned with one of the
shear eigendirections. The latter can be Fermi-transported
in each of these cases, e.g., ${\cal H}_2={\cal
H}_3=0$, as it follows  from (\ref{eq:alcon}) in the non degenerate case.
Using the algebraic constraints (\ref{eq:qom}) one sees that now equations 
(\ref{eq:dsigdotp1})--(\ref{eq:dmag3dot1}) are
reduced to a planar system:
\bea
\l{eq:plan1}
 \parb_{0} \Sigma_+ & = & (2-\Sigma_+)\,(1-\Sigma_+{}^2 -\Sigma_-{}^2)\;, \\
\l{eq:plan2}
 \parb_{0} \Sigma_- & = & - \,\Sigma_-\,(1-\Sigma_+{}^2 -\Sigma_-{}^2)\;,
\eea
where from  (\ref{eq:qom}) one has $1-\Sigma_+{}^2-\Sigma_-{}^2\geq 0$.
The system above is the heart of the dynamics of the more general case
when the magnetic field is not aligned; it 
has  one isolated stationary point at $\Sigma_+=2$,
$\Sigma_-=0$ (outside the physical phase space), and is also
stationary on the whole Kasner ring
$\Sigma_+{}^2+\Sigma_-{}^2=1$. The eigenvalues of the Jacobian at points
on the Kasner  ring are $\lambda_1=0$ and $\lambda_2=-\,4\,\Sigma_+
+2$. The first vanishes because there is no tangential motion along
the ring, the second is non-negative, $\lambda_2\geq 0$, for
$\Sigma_+\leq 1/2$. Thus, one third of the Kasner ring is
asymptotically stable, because of the aligned magnetic field.
These local properties of the Kasner ring are better understood
noticing that the planar system above admits  the  explicit solutions
\be
\Sigma_-=C\,(\Sigma_+ -2)\;, ~~~~-\,1/\sqrt{3}\leq C\leq 1/\sqrt{3}\;,
\ee
which are heteroclinic orbits joining pairs of points on the ring (they
are straight lines passing through the point $\Sigma_+=2$, $\Sigma_-=0$).
Since the magnetic field can be aligned with any of the three spatial
directions, there are two other stationary points of the system 
(\ref{eq:dsigdotp1})--(\ref{eq:dmag3dot1})
 in the $(\Sigma_+,\Sigma_-)$-plane, located at $\Sigma_+=-\,1$,
$\Sigma_-=\pm\,\sqrt{3}$. Passing through these two points, 
two other sets of heteroclinic orbits arise. The
overall effect of this structure in the physical phase space is that the
Kasner ring becomes  completely unstable, as each of its points becomes
a saddle.

\section{Conclusions}
The phase space structure in this example is reminiscent of the
Bianchi Type--IX dynamics, in that the Kasner ring no longer is a source
at any point. Thus, the approach towards the initial singularity
becomes more complicated and probably leads to an ``oscillatory''
regime. Unfortunately, it is harder to obtain a detailed understanding of 
this initial behaviour in the present situation because of the 
existence of the constraints (\ref{eq:alcon}). We do not
want to address the question if this behaviour is chaotic
or not. This theme has been debated for a
long time in the context of the Bianchi Type--IX models (see the
contribution of Di Bari {\it et al\/} in this volume). Clearly,
further work is needed to clarify the dynamics in the
vicinity of the singularity of the Bianchi Type--I models
with a magnetic field. In particular, a numerical
investigation would be desirable.

The Kasner ring is a source for 
Bianchi Type--I dust spacetimes with no
magnetic field.\cite{bi:WH} 
We have seen that this is no longer the case when magnetic fields are 
included. Indeed, the important point is that
a magnetic field completely changes the structure of the
phase space in the vicinity of the initial singularity, 
and as one gets closer to the singularity this effect becomes more and more
relevant. Hence, contrary to some folklore, matter fields can
strongly influence the dynamics close to the initial
singularity.\\

\noindent
{\bf Acknowledgements:}
MB would like to thank ICTP and INFN for financial support, and
SISSA for hospitality.

\end{document}